\documentclass{article}

\usepackage{arxiv}

\usepackage[utf8]{inputenc} % allow utf-8 input
\usepackage[T1]{fontenc}    % use 8-bit T1 fonts
\usepackage{hyperref}       % hyperlinks
\usepackage{url}            % simple URL typesetting
\usepackage{booktabs}       % professional-quality tables
\usepackage{amsfonts}       % blackboard math symbols
\usepackage{nicefrac}       % compact symbols for 1/2, etc.
\usepackage{microtype}      % microtypography
\usepackage{lipsum}
\usepackage{graphicx}
\usepackage{subcaption}
\graphicspath{ {./images/} }

\title{momenTUM: A Schema-Driven Platform For Designing, Deploying, and Exploring Ecological Momentary Assessment Studies}

\author{
Nataliia Petliak$^{1}$ \and
Blen Assefa$^{1,*}$ \and
Constantin Goeldel$^{1,*}$ \and
Marco Ma$^{1,*}$ \and
Anna Biller$^{1}$ \and
Lorna Caddick$^{5}$ \and
Raahat Manrai$^{5}$ \and
Daniel Smith$^{5}$ \and
Duncan Swiffen$^{5}$ \and
Manuel Spitschan$^{1,2,3,4,\dagger}$ \\
\\
$^{1}$TUM School of Medicine and Health, Technical University of Munich, Munich, Germany \\
$^{2}$Max Planck Institute for Biological Cybernetics, Translational Sensory \& Circadian Neuroscience, Tübingen, Germany \\
$^{3}$TUM Institute for Advanced Study (TUM-IAS), Technical University of Munich, Garching, Germany \\
$^{4}$TUMCREATE Ltd., Singapore, Singapore \\
$^{5}$Centre for Clinical Brain Sciences, The University of Edinburgh, United Kingdom \\
$^{*}$Equal contribution \\
$^{\dagger}$Corresponding author
}
\begin{document}
\maketitle
\begin{abstract}
Ecological momentary assessment (EMA) is widely used to collect repeated self-report data in participants' everyday lives using mobile devices. EMA studies often involve multiple questionnaires, flexible schedules, and longitudinal data collection, requiring reliable systems for study setup, deployment, monitoring, and data management. Existing workflows are often fragmented across tools, making studies difficult to reproduce and maintain.

We present momenTUM, an open-source platform for designing, deploying, and managing mobile EMA studies. Its central principle is that a structured study specification serves as the shared representation across the full workflow. The same specification supports authoring in the Study Designer, execution in the participant-facing mobile application, backend synchronization and storage, REDCap-linked data handling, and researcher monitoring. This makes protocols reusable, inspectable, and consistent across system components without requiring study-specific app implementations.

momenTUM integrates with REDCap to automate project setup and synchronize responses. Its authenticated dashboard provides tabular and calendar-based views, filtering by study components, and visual summaries. The platform has been deployed in real-world studies, including AMBIENT-BD, which examines mood, sleep, and circadian rhythms in bipolar disorder, and the EcoSleep cohort study. We also describe an exploratory LLM-assisted extension that generates draft structured study specifications for researcher review.

These deployments show that momenTUM can support complex ambulatory assessment protocols while reducing technical overhead and enabling reproducible, reusable, and extensible EMA workflows.

\end{abstract}

% keywords can be removed

\section{Introduction}

Ecological momentary assessment (EMA) is now widely used in research to collect repeated self-reports in participants’ everyday lives. It is especially useful in settings where the phenomena of interest change over time and are difficult to capture accurately through one-off or retrospective measures, such as mood, sleep, symptoms, behaviour, and contextual factors. As smartphones have become a standard part of daily life, they have also become a practical medium for delivering EMA protocols and collecting responses over extended periods.

At the same time, running EMA studies in practice is often more complicated than the basic idea suggests. Even relatively simple studies may involve multiple questionnaires, repeated notifications, branching logic, longitudinal schedules, participant management, and researcher-facing data inspection. In many cases, these functions are distributed across different tools, which makes workflows harder to maintain, update, and reproduce. As a result, the challenge is not only methodological, but also infrastructural: researchers need systems that can support study design, deployment, and downstream data access in a consistent way.

A central aim of momenTUM is to address this practical fragmentation by treating the study specification itself as the core object of the system. Rather than handling study design, app behaviour, and data interpretation as loosely connected steps, momenTUM uses a structured study specification that is shared across components. The same study representation is used to support authoring in the Study Designer, execution in the participant-facing mobile app, and inspection of collected data in the researcher dashboard. This makes the workflow more consistent and reusable, and helps preserve traceability across the study lifecycle.

In this paper, we describe momenTUM as an end-to-end EMA platform and focus in particular on the role of the structured study specification as the shared representation across study authoring, mobile execution, backend synchronization, REDCap-linked data handling, and researcher-side monitoring. We also describe an exploratory extension for LLM-assisted study generation, intended to support the creation of draft study specifications for further review, and discuss adaptive EMA as an important future direction for the platform.

% Ecological Momentary Assessment (EMA) has become a widely used methodology for studying behaviour, affect, symptoms, and contextual factors in daily life....

% Problem: actual workflow may be messy, not reproducible, different tools - fragmentation
% Hard to update studies, reproduce
% Also technical barrier 

% momenTUM:
% - everything is based on a structured study specification (JSON schema)

% - same thing is used for building a study, running it and interpreting data

% - consistency, reproducibility etc

% LLM idea
% helps with reducing barrier, faster prototyping

% future adaptive EMA: studies change based on answers, rule-based

% Contributions of paper:
% - describe the system end-to-end

% - schema as central concept

% - LLM prototype

\section{Related Work}
\label{sec:headings}

% EMA in general

% existing EMA platforms

EMA refers to the repeated sampling of participants’ current experiences, behaviours, and contexts in real time and in natural environments. This makes it particularly useful for studying phenomena that fluctuate over time, such as affect, symptoms, sleep, behaviour, and situational influences, while reducing reliance on retrospective recall. More broadly, EMA belongs to the family of ambulatory assessment methods that aim to capture ecologically valid data from everyday life and to characterize within-person variation as it unfolds over time. With the widespread adoption of smartphones, EMA has become substantially easier to deploy in practice, enabling repeated assessments at scale across longer study periods and more diverse settings. At the same time, this shift has made EMA increasingly dependent on software infrastructure for scheduling, delivery, response capture, and downstream data handling. \cite{shiffman_ecological_2008, trull_ambulatory_2013}

A number of digital platforms have been developed to support ecological momentary assessment and related mobile intervention workflows. For example, MindLogger \cite{klein_remote_2021} provides an open and configurable platform for creating and scheduling mobile assessments, tasks, and interventions without requiring programming expertise, with support for administrator-side study setup and participant-facing delivery through iOS and Android applications. Similarly, m-Path \cite{mestdagh_m-path_2023} offers a highly tailorable framework for implementing smartphone-based EMA and ecological momentary intervention in both research and clinical contexts. Other systems have moved further toward context-aware and reactive functionality; for instance, AwarNS \cite{gonzalez-perez_awarns_2023} is designed as a framework for developing mobile health and mental health applications that incorporate contextual information and reactive logic. Together, these systems demonstrate that smartphone-based EMA is now supported by a diverse software landscape that spans configurable assessment platforms, intervention-oriented systems, and context-aware mobile frameworks.  

While these platforms demonstrate that smartphone-based EMA is already supported by a mature and diverse software ecosystem, they also emphasize different parts of the workflow, such as configurable assessment delivery, intervention support, or context-aware mobile behaviour. momenTUM is positioned within this landscape as an open-source research platform in which the study specification serves as the shared representation across study authoring, participant-facing execution, backend storage and synchronization, REDCap-linked data handling, and researcher-side monitoring. The contribution of momenTUM therefore lies less in introducing a new EMA modality than in providing a coherent end-to-end workflow in which study protocols remain reusable, inspectable, and operationally consistent across these components.

\section{Design Goals and Rationale}

momenTUM was developed to address recurring practical challenges in implementing EMA studies with mobile devices in research settings. Although smartphone-based data collection is now common, the surrounding workflow often remains fragmented across multiple tools for study setup, participant delivery, data synchronization, and researcher-side inspection. This fragmentation makes study protocols harder to maintain, update, and reuse across repeated study periods or related projects. 

The design of momenTUM was guided by several goals. First, study logic should be represented explicitly rather than being distributed implicitly across application code, external spreadsheets, and platform-specific settings. This motivated the use of a structured study specification as the central representation of the protocol.

Second, the same study representation should be usable across the main stages of the workflow. In momenTUM, the study specification is used for authoring in the Study Designer, execution in the mobile application, and interpretation in the researcher dashboard. This helps reduce inconsistencies between what researchers configure, what participants see, and what is later inspected during data collection.

Third, the system should support reuse across study periods and across different research contexts. Rather than developing a separate mobile application for each study instance, the goal was to separate study content and scheduling logic from the app implementation itself. This makes it possible to adapt or extend studies by modifying their structured configuration while reusing the same app and backend infrastructure.

Fourth, the platform should support practical researcher workflows beyond questionnaire delivery alone. In the current system, this includes REDCap integration for project setup and response synchronization, as well as dashboard-based inspection of collected responses, enrollment, and adherence. These elements are important in practice because many EMA studies require not only delivery of scheduled tasks, but also ongoing operational monitoring during data collection.

Finally, the platform was designed with extensibility in mind. In addition to supporting current EMA workflows, the architecture is intended to support further development in areas such as natural-language-assisted study authoring, more flexible notification behavior, and adaptive EMA logic.

\section{System Overview}

momenTUM consists of four main components: a web-based Study Designer, a participant-facing mobile application for iOS and Android, backend services for study storage and synchronization, and a researcher dashboard for inspecting collected data. These components are connected through a shared structured study specification, which acts as the common representation of the study across the system.

\begin{figure*}[!t]
    \centering
    \includegraphics[width=\textwidth]{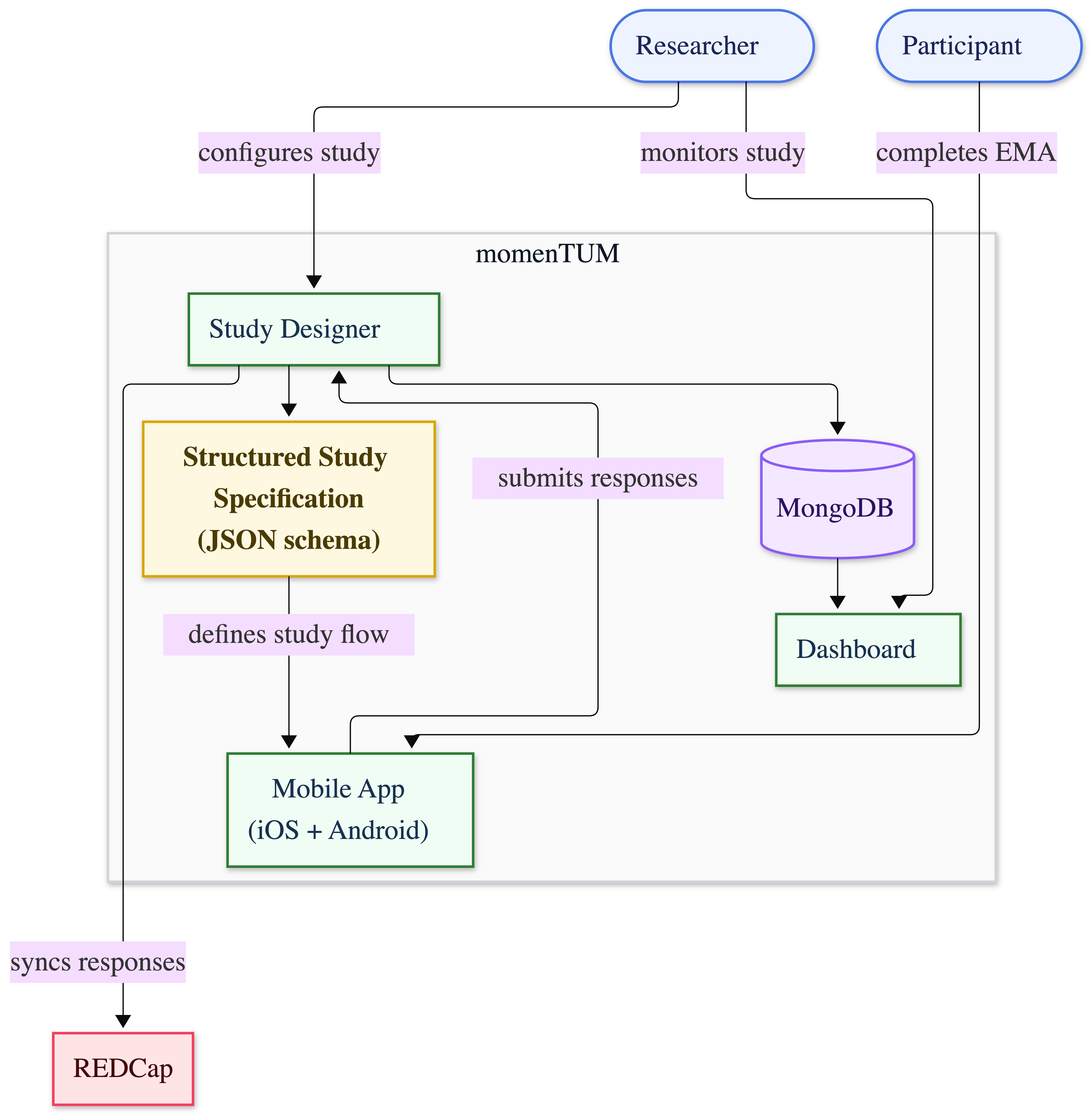}
    \caption{Overview of the main momenTUM workflow. Researchers configure studies in the Study Designer and monitor data collection through the dashboard, while participants complete EMA tasks in the mobile application. A shared structured study specification connects study authoring, mobile delivery, backend storage in MongoDB, and synchronization with REDCap.}
    \label{fig:architecture}
\end{figure*}

Figure~\ref{fig:architecture} illustrates the main workflow. Researchers use the Study Designer to configure study structure, scheduling, and content. The resulting study specification is stored in the backend and made available to the mobile application, which renders the participant-facing study flow and generates tasks according to the configured scheduling rules. Participant responses are stored in the backend database and can be synchronized with linked REDCap projects. The researcher dashboard provides access to collected data and supports filtering, calendar-based inspection, and adherence monitoring.

The Study Designer is the main authoring interface of the platform. It allows researchers to define study-level properties, modules, scheduling behavior, question types, and deployment-related actions such as study upload and QR-code-based participant access. The participant-facing mobile application delivers study content through a shared cross-platform app rather than requiring a separate app build for each study. The dashboard provides a monitoring layer on top of the collected data and supports both participant-level inspection and higher-level study oversight.

At the data layer, the system stores structured study specifications and collected responses in MongoDB and synchronizes response data with REDCap through the REDCap API. This supports both operational workflows during data collection and downstream integration with established research data management infrastructure.

\subsection{Implementation}

From an implementation perspective, momenTUM combines web, mobile, backend, and infrastructure components that together support study authoring, deployment, synchronization, and monitoring. The participant-facing mobile application builds on the earlier schema-based app foundation \cite{shatte_schema_2020}, while the surrounding system extends this mobile layer through study authoring, REDCap-linked synchronization, and dashboard-based monitoring. Table~\ref{tab:implementation} summarizes the main implementation components.

\begin{table}[t]
\centering
\caption{Main implementation components of momenTUM.}
\label{tab:implementation}
\begin{tabular}{lll}
\toprule
Subsystem & Role & Implementation \\
\midrule
Study authoring subsystem & Study configuration, upload, and management & Web-based frontend + backend API \\
Mobile subsystem & Participant-facing EMA delivery & Ionic-based app building on schema \\
Dashboard subsystem & Researcher-facing data inspection and monitoring & Web-based frontend + backend API \\
Data and synchronization layer & Study storage, logs, and REDCap synchronization & FastAPI (Python), MongoDB \\
Infrastructure & Deployment and hosting & Dockerized services on LRZ servers \\
\bottomrule
\end{tabular}
\end{table}

\section{Reproducible Study Specification}

A central design choice in momenTUM is that each study is represented as a structured JSON-based study specification rather than being distributed implicitly across application code, ad hoc configuration files, and platform-specific settings. This specification serves as the shared representation of the protocol across the main components of the system and is used to support study authoring, mobile execution, backend synchronization, and researcher-facing data inspection.

At a high level, the study specification contains two main layers: study-level properties and a list of modules. Study-level properties capture metadata and participant-facing information such as the study identifier, study name, instructions, support links, ethics information, and server endpoints. Modules represent the concrete units of study activity and contain the configuration needed for scheduling, participant interaction, and downstream visualization. In the current system, modules can include, for example, questionnaire-based survey tasks and psychomotor vigilance task (PVT) modules.

Within each module, the specification stores multiple kinds of logic. First, it defines scheduling and alert behavior, including whether notifications are scheduled at absolute calendar times or relative to participant enrollment. It also captures repeat frequency, interval, randomization of notification timing, timeout behavior, sticky notifications, and additional times of day. Second, it defines the participant-facing content of the module, including sections, questions, and type-specific settings. The current study representation supports multiple question types, including text, numeric text, yes/no, datetime, slider, multiple-choice, media, instruction, and photo-based items. Third, it captures behavioral rules such as whether a question is required, simple visibility logic, module unlocking dependencies, and graph-related settings for progress visualization.

At the implementation level, this specification is backed by an explicit JSON schema that defines required fields, nested object structure, identifier constraints, and type-specific validation rules for different module and question configurations. 

In this context, reproducibility means that the same study protocol can be represented and executed consistently across repeated deployments and across system components. More concretely, the structured study specification preserves the schedule, module structure, question definitions, response options, conditional logic, identifiers and variable mapping, and the participant-facing app behaviour derived from these elements. Because these aspects are represented explicitly in a shared machine-readable form, the protocol can be deployed, synchronized, and inspected consistently across the platform.

A simplified excerpt of the study specification is shown below:

\begin{verbatim}
{
  "properties": {
    "study_name": "Example study",
    "study_id": "example_study"
  },
  "modules": [
    {
      "id": "mood_diary",
      "name": "Mood Diary",
      "condition": "*",
      "alerts": {
        "scheduleMode": "relative",
        "repeat": "daily",
        "interval": 1,
        "offsetDays": 0,
        "offsetTime": "08:00:00"
      },
      "params": {
        "type": "survey",
        "submit_text": "Submit",
        "sections": [
          {
            "name": "Mood",
            "questions": [
              {
                "type": "slider",
                "text": "How is your mood right now?",
                "required": true,
                "min": 0,
                "max": 10
              }
            ]
          }
        ]
      }
    }
  ]
}
\end{verbatim}

This structured JSON representation plays an important role in keeping the workflow consistent across the platform. In the Study Designer, the specification is edited through a graphical interface and stored in an explicit machine-readable form. In the mobile application, the same representation is interpreted to render study content and generate participant tasks according to the configured scheduling rules. In the backend and dashboard, study structure and collected responses can be linked back to the same underlying identifiers and module definitions. Because the study logic is represented explicitly rather than embedded separately in multiple components, the specification helps preserve traceability between what was designed, what was delivered to participants, and what is later synchronized, stored, and inspected by researchers.

The specification also supports reuse across study periods and related study contexts. Rather than requiring the development of a separate mobile application for each study instance, a new study version can be created by modifying the JSON study specification while reusing the same app and backend infrastructure. In practice, this is particularly useful in longitudinal and repeated-burst study designs, where study content, timing, or module composition may need to be adjusted between periods without changing the underlying technical platform.

% Figure~\ref{fig:schema} summarizes the main structure of the study specification used across the platform.

From a methodological perspective, the main value of the study specification is therefore not only that it provides a machine-readable configuration format, but that it acts as the operational definition of the study across the system. It describes not just what questions exist, but how the study behaves over time: how participants enter it, when tasks appear, what content is shown, and how responses are linked back to a structured representation that can be synchronized and inspected consistently.

% \begin{figure*}[!t]
%     \centering
%     \includegraphics[width=\textwidth]{schema.png}
%     \caption{Structure of the momenTUM study specification. A study contains study-level properties and a list of modules. Each module combines scheduling, conditions, and content parameters. Survey modules contain sections, which in turn contain questions with shared and type-specific fields.}
%     \label{fig:schema}
% \end{figure*}

% \begin{figure*}[!t]
%     \centering

%     \begin{subfigure}[t]{0.92\textwidth}
%         \centering
%         \includegraphics[width=\textwidth]{schema_2a.png}
%         \caption{}
%     \end{subfigure}

%     \vspace{0.6em}

%     \begin{subfigure}[t]{0.4\textwidth}
%         \centering
%         \includegraphics[width=\textwidth]{schema_2b1.png}
%         \caption{}
%     \end{subfigure}

%     \vspace{0.5em}

%     \begin{subfigure}[t]{0.5\textwidth}
%         \centering
%         \includegraphics[width=\textwidth]{schema_2b2.png}
%         \caption{}
%     \end{subfigure}

%     \vspace{0.5em}

%     \begin{subfigure}[t]{0.4\textwidth}
%         \centering
%         \includegraphics[width=\textwidth]{schema_2b3.png}
%         \caption{}
%     \end{subfigure}

%     \caption{Structure of the momenTUM study specification. (a) High-level hierarchy of study, module, survey, section, and question objects. (b) Examples of study-level property fields. (c) Alert and scheduling fields. (d) Question-level fields, including shared, visibility-related, and type-specific settings.}
%     \label{fig:schema}
% \end{figure*}

\section{Study Authoring with the Study Designer}

\begin{figure*}[!t]
    \centering
    \includegraphics[width=0.85\textwidth]{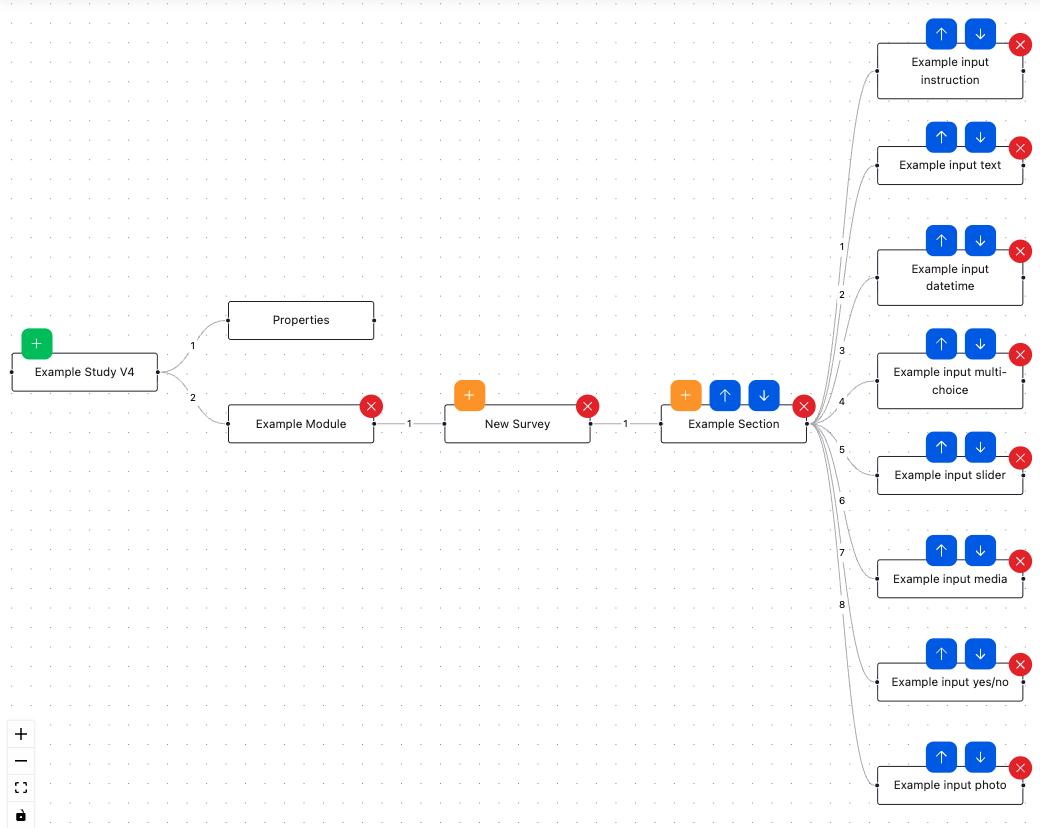}
    \vspace{0.5em}

    \includegraphics[width=0.4\textwidth]{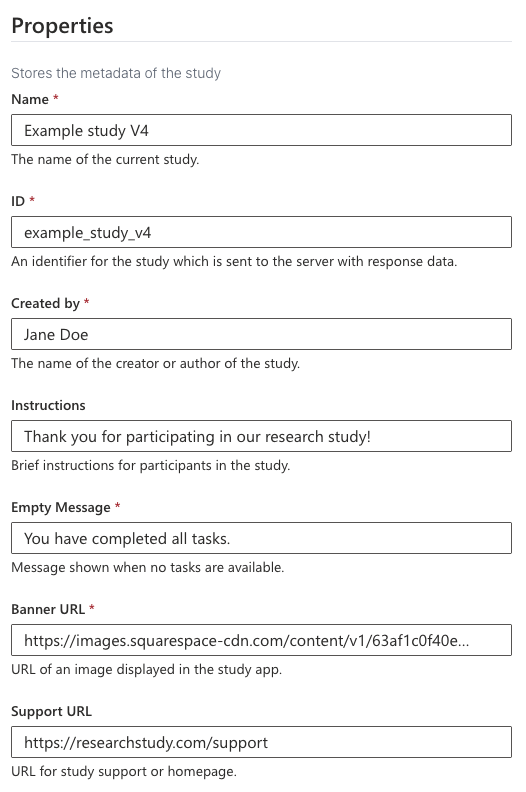}
    \caption{Example views from the momenTUM Study Designer. Top: hierarchical study structure showing the organization of a study into properties, modules, sections, and different question types. Bottom: study-level properties used to define identifiers, participant-facing instructions, and support-related metadata.}
    \label{fig:designer_overview}
\end{figure*}

The Study Designer provides the main authoring interface for creating structured EMA studies in momenTUM. Rather than editing raw JSON directly, researchers configure studies through a graphical interface that maps onto the shared study specification used throughout the platform. Figure~\ref{fig:designer_overview} shows example views from the Designer, including the hierarchical study layout and the study-level properties panel.

In the Designer, researchers define study-level metadata together with the modules that make up the study protocol. Each module corresponds to a concrete unit of study activity and can be configured with its own scheduling behavior, participant-facing content, and supporting parameters. This allows studies to be constructed as collections of separately configurable units rather than as single monolithic questionnaires.

The Designer exposes module-level scheduling controls that map directly onto the shared study specification, including absolute and enrollment-relative timing, repeat behavior, additional times of day, message text, and notification-related settings such as sticky prompts, randomized timing, and timeout behavior. In this way, both simple repeated questionnaires and more structured longitudinal task schedules can be authored within the same interface.

Within survey modules, content is organized into sections containing typed questions. The Designer supports shared question properties such as text, required status, and optional visibility logic, together with type-specific configuration for supported response formats including text, yes/no, datetime, multiple-choice, media, slider, instruction, and photo-based items. This preserves consistency between what is authored in the interface and what is later rendered in the participant-facing application.

The Study Designer therefore acts not only as a user interface for entering study content, but as the main authoring layer for constructing, validating, and revising the structured study specification that is later deployed, synchronized, and inspected across the rest of the platform.

\section{Mobile Deployment and Participant Experience}

The participant-facing mobile application provides the execution layer of momenTUM and is responsible for enrolling participants into studies, rendering study content, scheduling and displaying tasks, collecting responses, and returning completed data to the backend. The app is implemented as a shared cross-platform mobile application for iOS and Android and builds on the earlier schema-based app foundation \cite{shatte_schema_2020}. Within momenTUM, this mobile layer is extended and integrated into a broader workflow that includes study authoring, backend synchronization, REDCap integration, and researcher-facing monitoring.

A central property of the mobile subsystem is that it does not require a separate app build for each individual study. Instead, the app interprets the structured study specification produced in the Study Designer and uses it to render participant-facing study flows dynamically. This means that study content, module composition, scheduling behavior, and question structure are configured at the level of the study specification rather than being hard-coded into the application itself. As a result, new studies and revised study versions can be deployed through configuration and synchronization while reusing the same mobile infrastructure.

Participants can enroll into a study through multiple entry points, including QR-code scanning, direct URL-based enrollment, and manual study identifier entry. After enrollment, the app retrieves the study specification from the backend and generates participant tasks according to the scheduling logic defined in the study specification. The home screen provides access to currently available tasks and recently completed modules, while the progress view supports lightweight participant-facing feedback such as graph-based summaries for modules where this has been configured.

\begin{figure*}[t]
    \centering

    \begin{subfigure}[t]{0.26\textwidth}
        \centering
        \fbox{\includegraphics[width=0.95\linewidth]{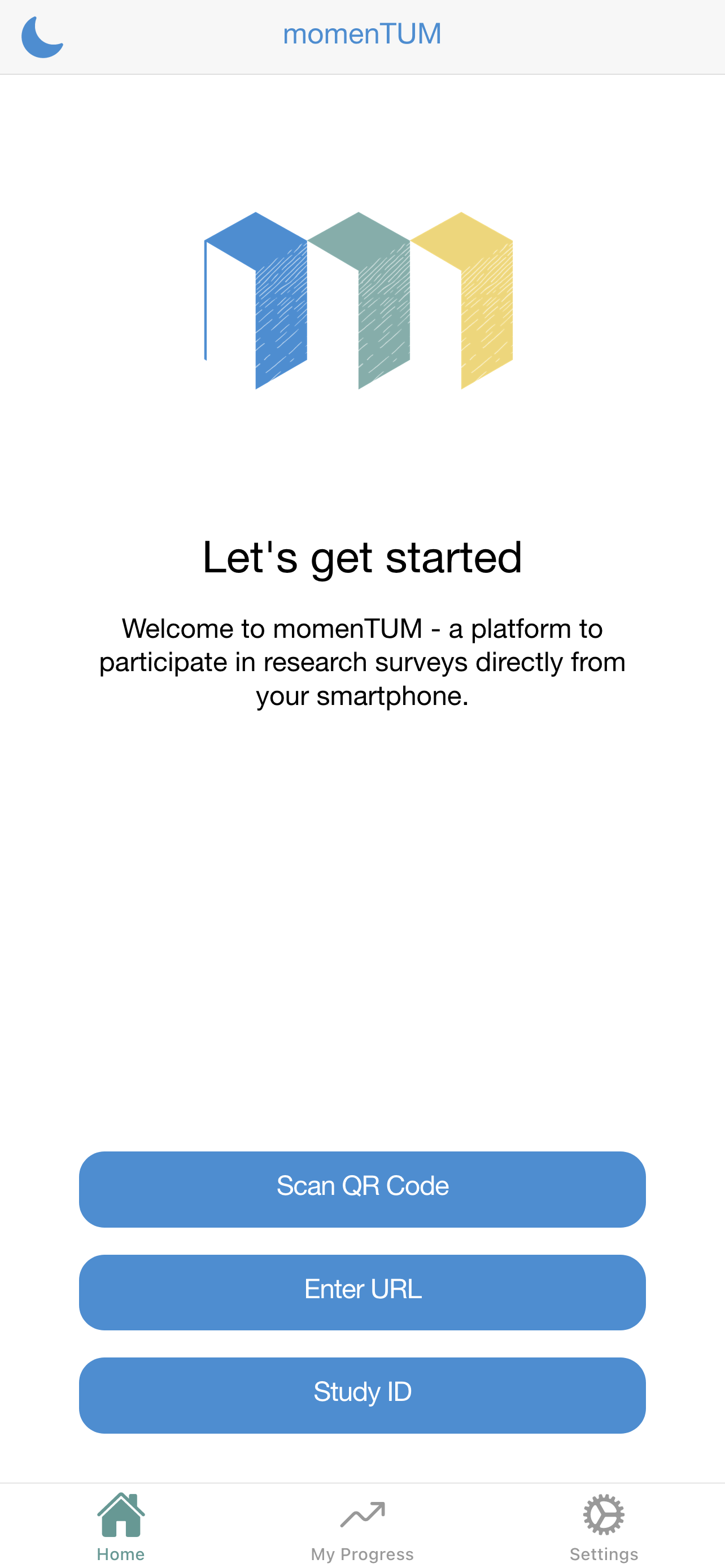}}
        \caption{Study enrolment}
    \end{subfigure}
    \hfill
    \begin{subfigure}[t]{0.26\textwidth}
        \centering
        \fbox{\includegraphics[width=0.95\linewidth]{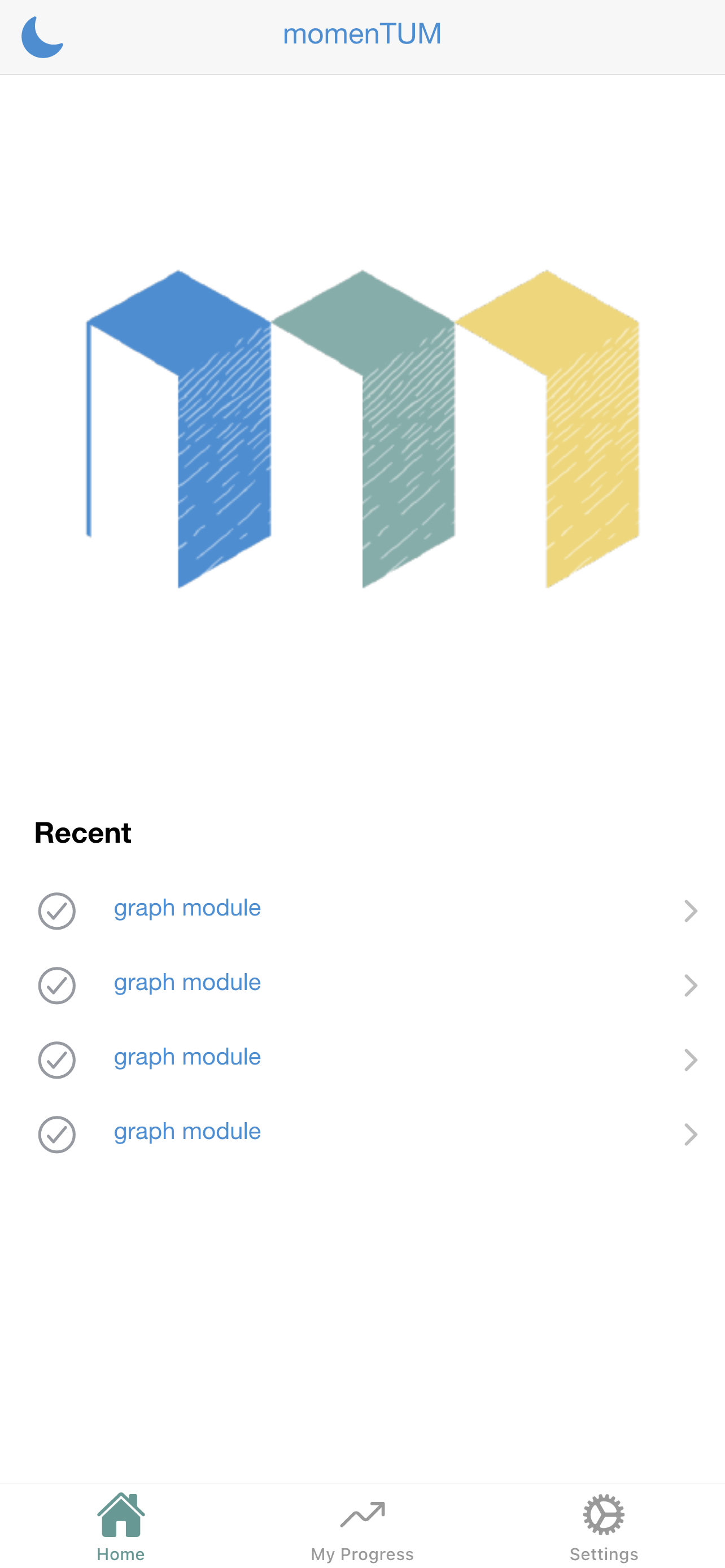}}
        \caption{Recent modules}
    \end{subfigure}
    \hfill
    \begin{subfigure}[t]{0.26\textwidth}
        \centering
        \fbox{\includegraphics[width=0.95\linewidth]{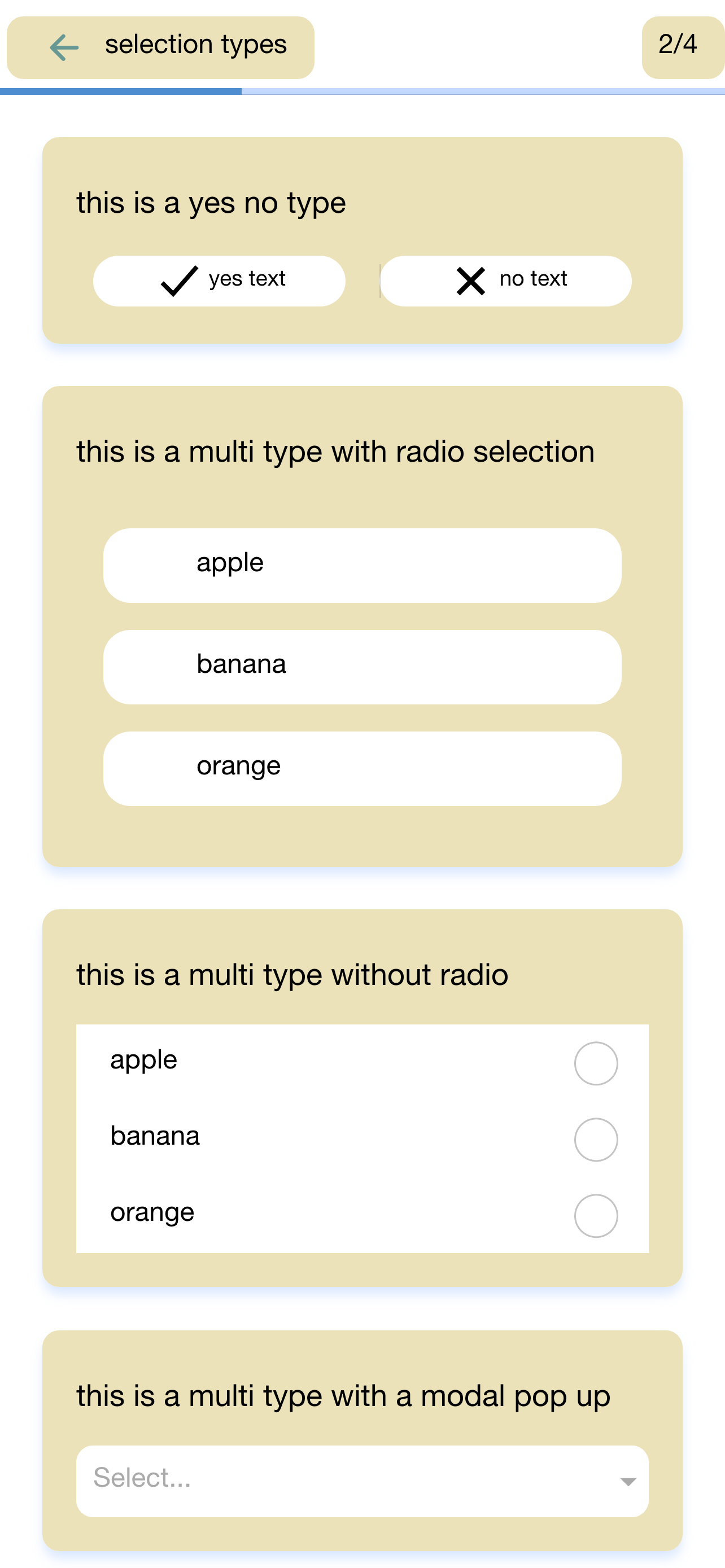}}
        \caption{Survey rendering}
    \end{subfigure}

    \vspace{0.8em}

    \begin{subfigure}[t]{0.26\textwidth}
        \centering
        \fbox{\includegraphics[width=0.95\linewidth]{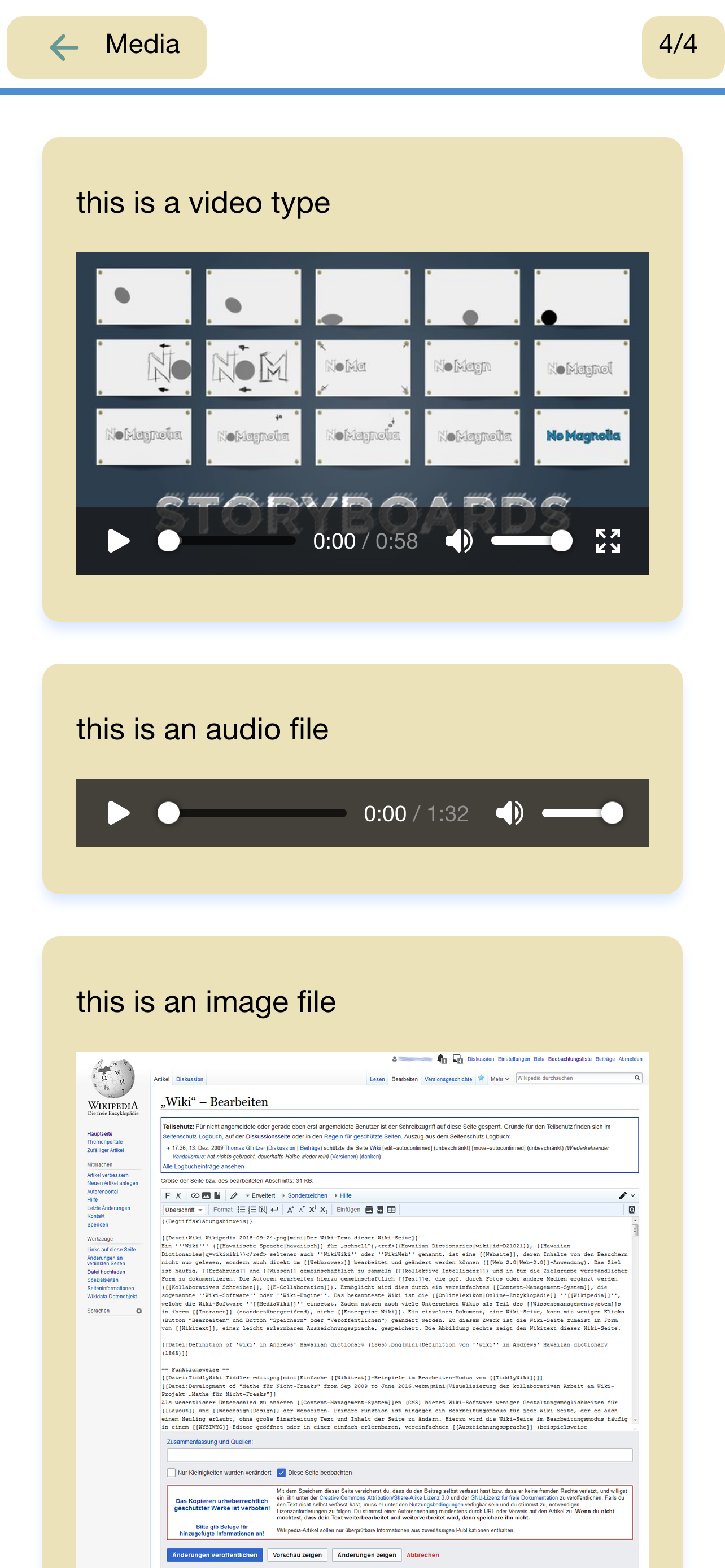}}
        \caption{Selection and media types}
    \end{subfigure}
    \hfill
    \begin{subfigure}[t]{0.26\textwidth}
        \centering
        \fbox{\includegraphics[width=0.95\linewidth]{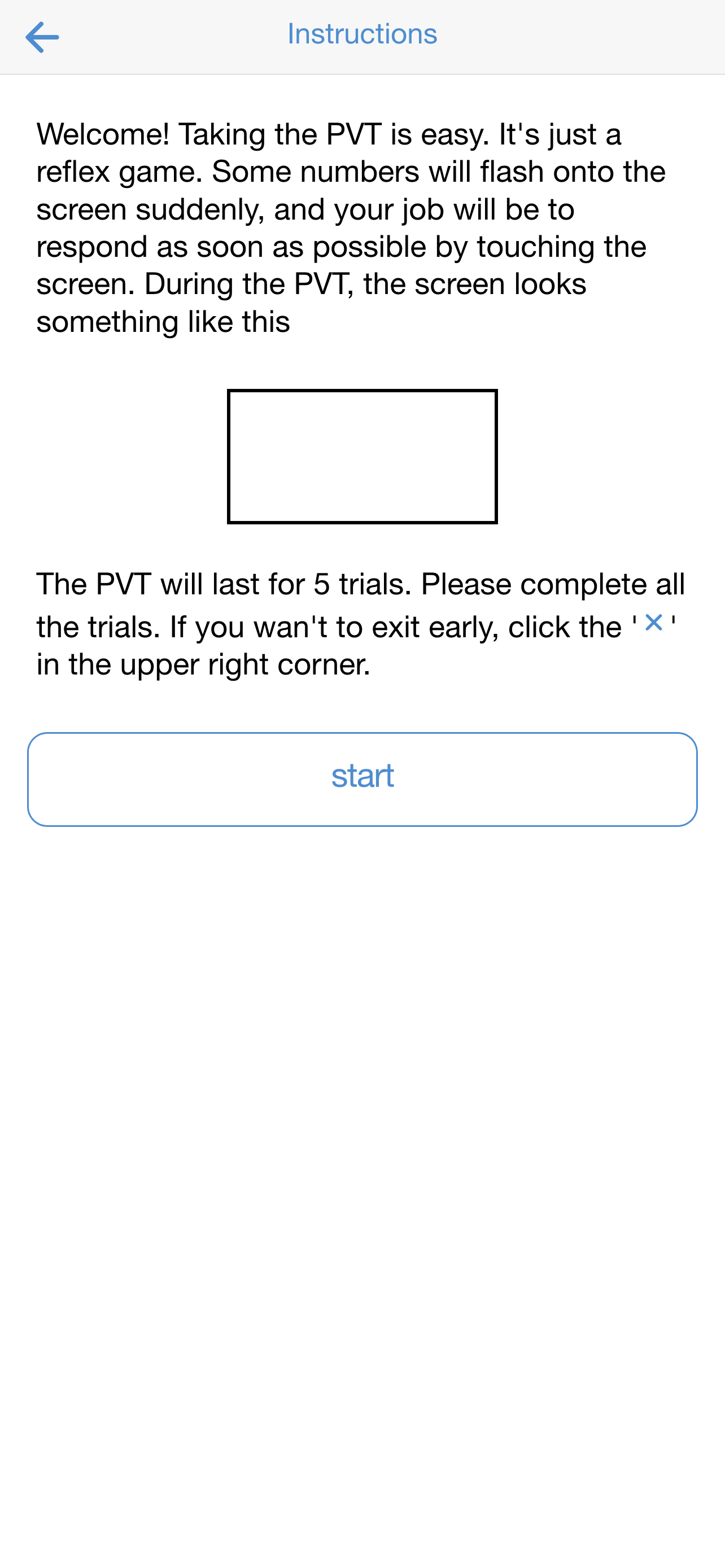}}
        \caption{PVT instructions}
    \end{subfigure}
    \hfill
    \begin{subfigure}[t]{0.26\textwidth}
        \centering
        \fbox{\includegraphics[width=0.95\linewidth]{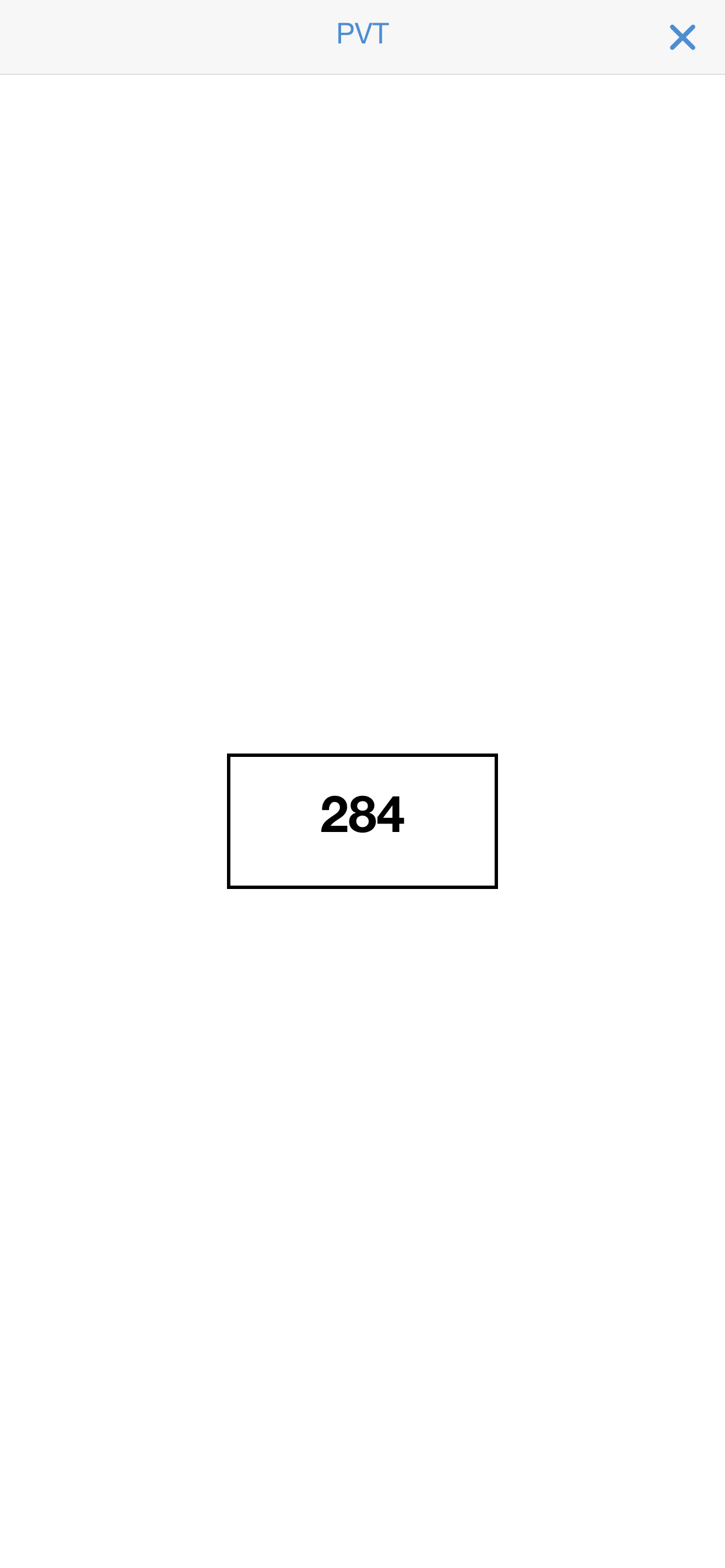}}
        \caption{PVT task}
    \end{subfigure}

    \caption{Example views from the momenTUM mobile application. The app supports study enrolment, display of available and completed modules, dynamic rendering of survey content from the shared study specification, multiple question and media types, and interactive task modules such as the psychomotor vigilance task (PVT).}
    \label{fig:mobile_app}
\end{figure*}

Study content is rendered directly from the structured representation of modules, sections, and questions. Survey modules are displayed as ordered sections with progress indicators and support the full set of configured question types, including short and long text input, numeric input, yes/no items, single- and multiple-selection items, datetime fields, sliders, instruction blocks, media elements, and photo-based questions. Because these question types are driven by the shared study specification, the same logical study structure used in authoring is preserved in the participant-facing interface. Figure~\ref{fig:mobile_app} shows example views from the app, including enrollment, task overview, questionnaire rendering, media support, progress visualization, and the psychomotor vigilance task interface.

Notification behaviour in momenTUM is derived from the study specification at the module level. The current system supports both absolute and enrollment-relative schedules, repeat intervals, additional times of day, randomized windows, sticky notifications, and timeout behaviour. After study enrolment, the mobile application retrieves the study specification and generates participant tasks and local notification timing accordingly. Actual delivery remains subject to operating-system behaviour and participant device settings, including cases such as Do Not Disturb mode, which can affect when prompts are seen in practice.

In addition to questionnaire-based modules, the mobile application also supports interactive task modules such as the psychomotor vigilance task (PVT). In these cases, the app renders a task-specific interface defined by the module type rather than a standard survey form. This makes it possible for the same study to combine repeated self-report questionnaires with other structured mobile assessment components within a single participant workflow.

Once a participant completes a task, responses are stored with the corresponding study and module identifiers and are transmitted to the backend for storage and downstream synchronization. This linkage is important because it preserves consistency between what was configured in the Study Designer, what was shown in the app, and what later appears in REDCap and the researcher dashboard. In this way, the mobile application functions not merely as a delivery interface, but as the operational execution layer of the shared study specification across real-world EMA data collection.

\section{Data Infrastructure and Synchronization}

The data infrastructure of momenTUM connects study authoring, mobile execution, persistent storage, and downstream research data management. At the backend level, the platform is implemented as a FastAPI service with separate routers for studies, responses, and REDCap-related operations, exposed through a versioned API. This backend acts as the coordination layer between the Study Designer, the participant-facing mobile application, the database, and external data management systems.

A central responsibility of the backend is the storage and retrieval of structured study specifications. Studies are stored in MongoDB together with timestamps and metadata, and can be retrieved either by internal object identifier or by study identifier. In practice, revised study deployments are preserved as distinct stored study specifications rather than overwriting prior configurations, which makes it possible to retrieve both the latest definition and earlier deployed versions. This is important because study specifications may evolve across repeated deployments or data collection periods, while the platform still needs to preserve a record of what was configured and when.

The same backend also receives completed participant responses from the mobile application. When a participant submits a module, the backend stores the response payload together with study and module identifiers, timing information, and device-related metadata. This creates a persistent record of what was completed in the app and provides the basis for later inspection in the dashboard and synchronization with external systems. Because responses are stored together with explicit study and module references, collected data remain linked to the structured study specification rather than existing as disconnected questionnaire entries.

A key feature of this layer is synchronization with REDCap. In the current system, the backend supports REDCap project setup and metadata import, including the creation of module-specific repeat instruments derived from the study structure. Completed mobile responses are then transformed into REDCap-compatible records and submitted through the REDCap API. In this mapping, response timing and module-level metadata are preserved alongside question-level values, allowing REDCap to serve as a downstream research data management layer while the mobile app and backend continue to operate through the shared study specification.

This synchronization layer is important for two reasons. First, it reduces manual overhead in configuring and maintaining parallel study structures across systems. Second, it helps keep study execution and stored research data aligned: the same identifiers used in the Study Designer and mobile app are reused in backend storage and REDCap export. In this way, the backend does not merely act as a transport layer, but as the component that maintains consistency between study specification, participant interaction, and downstream data representation across the platform.

momenTUM also includes basic access-control and data-handling measures appropriate for sensitive EMA workflows. Access to the researcher dashboard is authenticated, while the participant-facing mobile application does not require collection of direct personal identifiers such as names or email addresses. Instead, app data are linked through internally generated user identifiers, and studies may additionally collect researcher-assigned participant identifiers where cross-linkage to other study stages or external datasets is required. This separation helps support pseudonymised handling of participant data within the platform. The platform is hosted on institutional infrastructure, and study specifications and response data are stored and synchronized through the backend.

\section{Researcher Dashboard}

The researcher dashboard provides a web-based interface for inspecting and monitoring study data after collection. While the Study Designer is used to define study structure and the mobile application is used to execute participant-facing tasks, the dashboard serves a different purpose: it supports study-level review of submitted responses, participant progress, and derived visual summaries. In this way, it complements the rest of the platform by giving researchers direct access to collected data in a format that is easier to explore than raw database records or REDCap tables.

\begin{figure*}[t]
    \centering

    \begin{subfigure}[t]{0.5\textwidth}
        \centering
        \includegraphics[width=\linewidth]{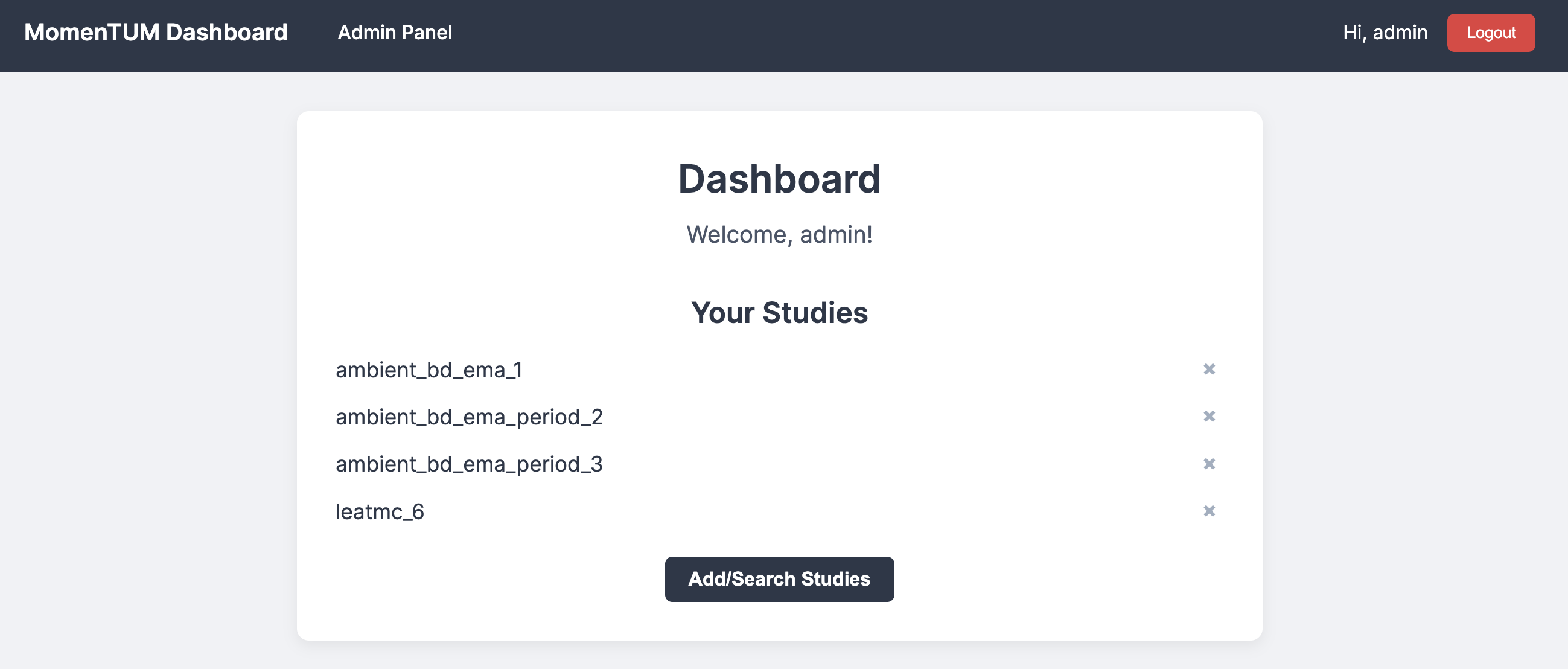}
        \caption{Study selection view}
    \end{subfigure}
    \hfill
    \begin{subfigure}[t]{0.5\textwidth}
        \centering
        \includegraphics[width=\linewidth]{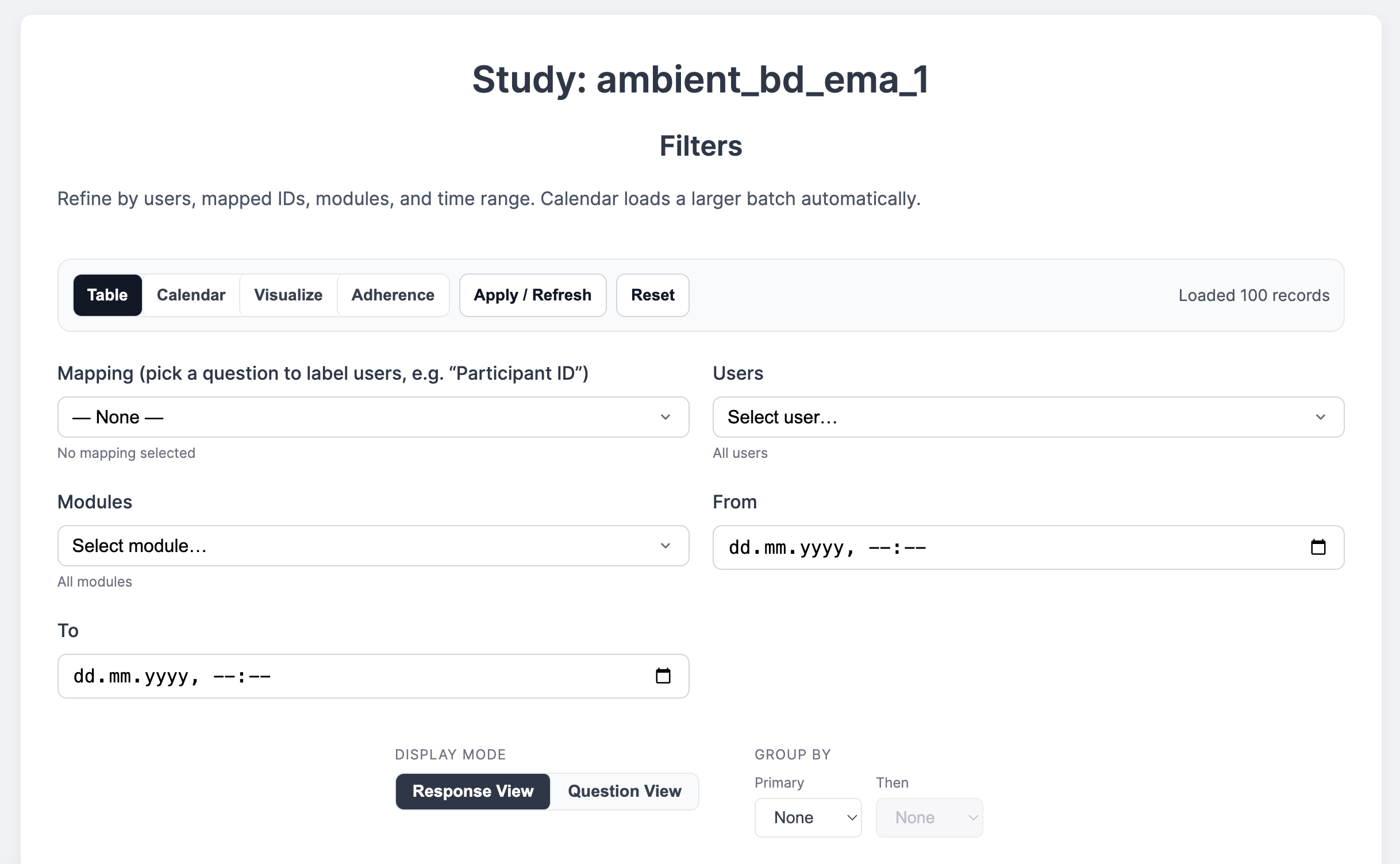}
        \caption{Study-level filters and display controls}
    \end{subfigure}
    \hfill
    \begin{subfigure}[t]{0.5\textwidth}
        \centering
        \includegraphics[width=\linewidth]{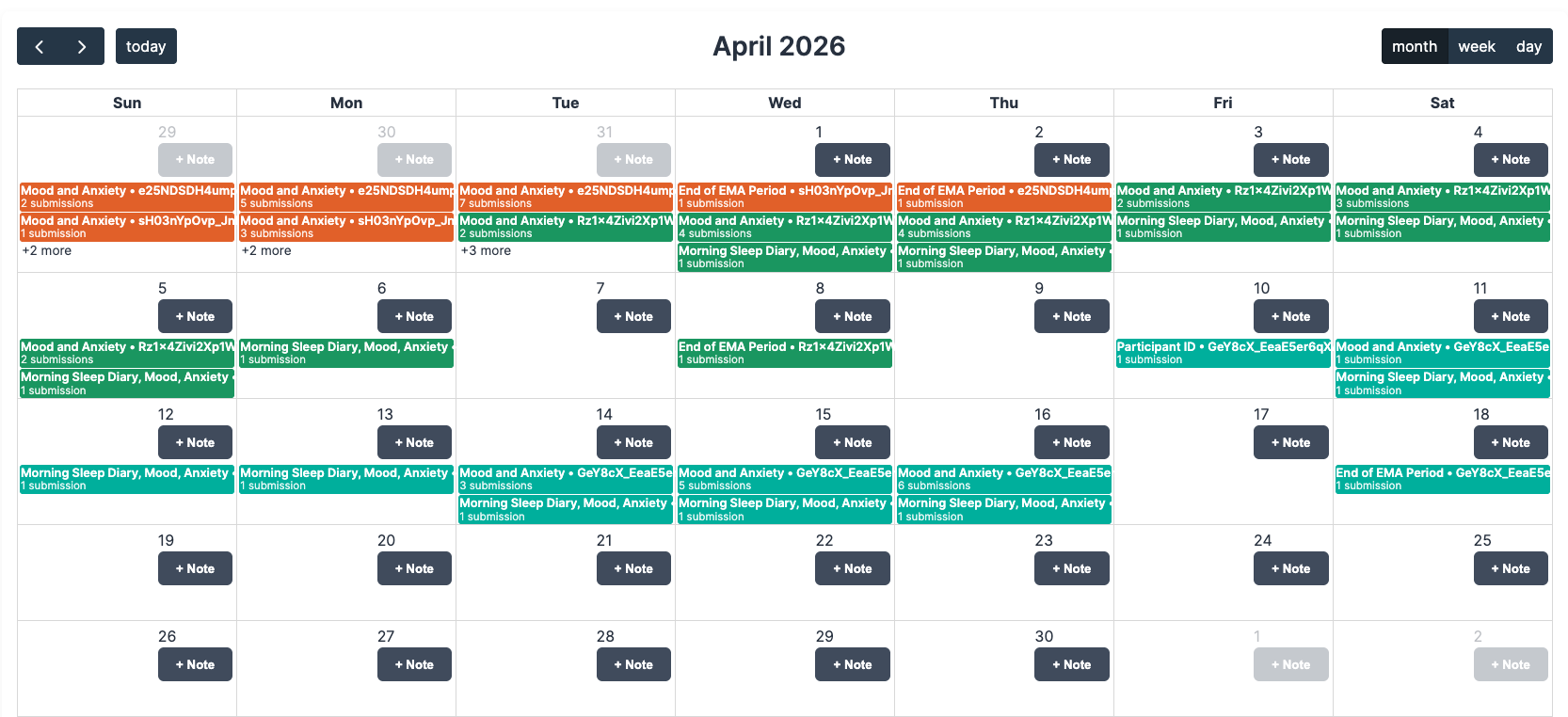}
        \caption{Calendar view of repeated submissions}
    \end{subfigure}

    \vspace{0.8em}

    \begin{subfigure}[t]{0.5\textwidth}
        \centering
        \includegraphics[width=\linewidth]{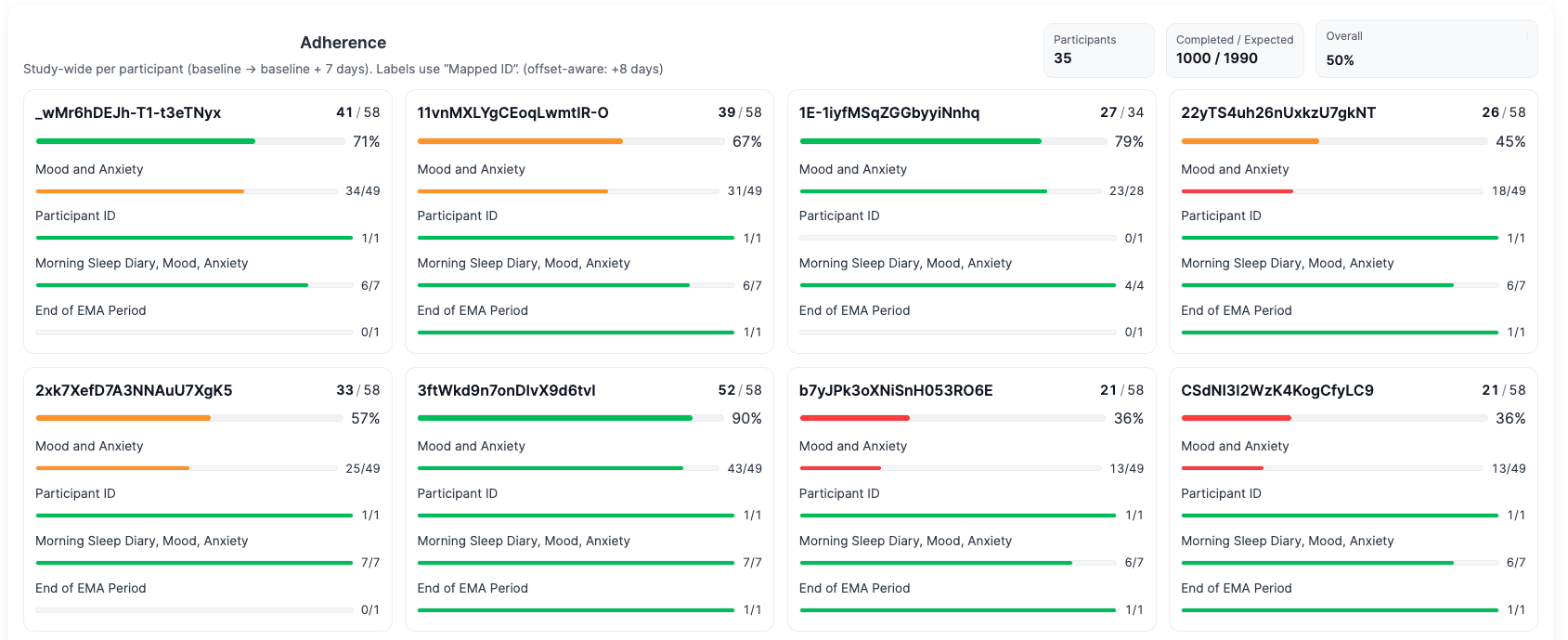}
        \caption{Participant-level adherence overview}
    \end{subfigure}
    \hfill
    \begin{subfigure}[t]{0.5\textwidth}
        \centering
        \includegraphics[width=\linewidth]{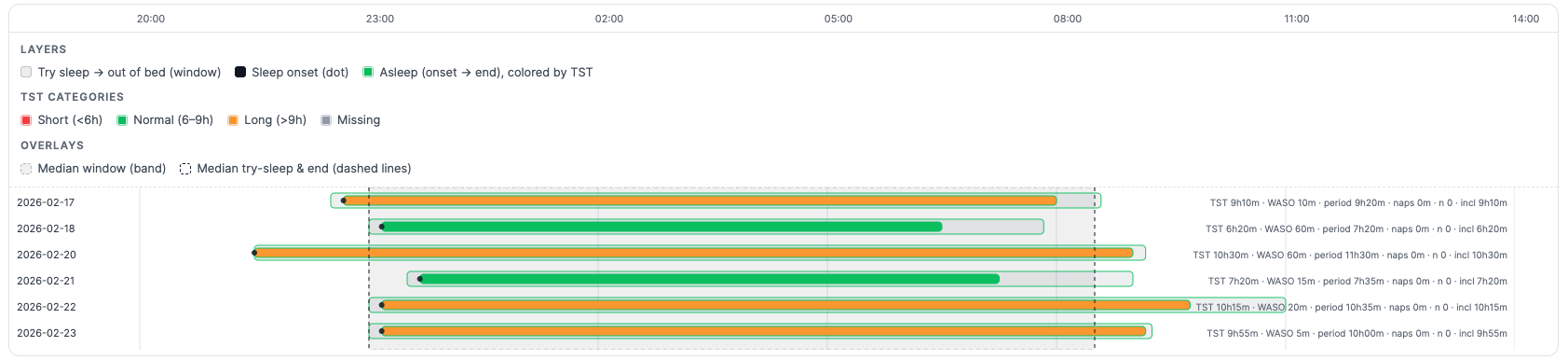}
        \caption{Study-specific visualization example}
    \end{subfigure}

    \caption{Example views from the momenTUM researcher dashboard. The dashboard supports study selection, filtering by participant labels and study metadata, calendar-based inspection of repeated submissions, adherence monitoring, and study-specific visual summaries derived from structured response data.}
    \label{fig:dashboard}
\end{figure*}

The dashboard is organized around study-specific views. After login, researchers can access the studies assigned to them and open an individual study workspace. Within each study, submitted responses can be explored through several coordinated views, including tabular inspection, calendar-based navigation, participant-level adherence summaries, and study-specific visualizations (Figure~\ref{fig:dashboard}). This structure supports both detailed review of individual submissions and broader monitoring of response patterns across time and participants.

A central design feature of the dashboard is its filtering and labeling layer. Researchers can refine displayed records by mapped participant identifier, internal user identifier, module, and time range. Because participant-facing studies may use internal app identifiers that are not directly meaningful to researchers, the dashboard supports mapping a selected question---for example, a participant ID field---to serve as the visible participant label. This makes the interface more interpretable in real study use while preserving the original identifiers in the underlying data model.

The dashboard supports multiple display modes for different inspection tasks. The table view is intended for direct browsing of response records and grouped question content. The calendar view places submitted modules on a time grid, which is useful for repeated-measures studies in which timing and spacing of responses matter. The adherence view summarizes completed versus expected responses at both participant and study level, making it easier to identify missing data, dropout patterns, or incomplete module sequences during ongoing data collection. In addition, the visualization layer supports study-specific summaries built on top of structured response data, such as sleep-related timelines and duration statistics for longitudinal monitoring.

This dashboard layer relies on the same study and question identifiers used throughout the rest of the platform. As a result, submitted mobile responses can be grouped, filtered, labeled, and visualized without losing their connection to the study specification that generated them. The dashboard therefore functions as the researcher-facing inspection layer of momenTUM, translating stored response data into views that support quality control, progress tracking, and early-stage exploratory review during active study deployment.

\section{LLM-Assisted Study Generation}

To reduce the manual effort required to create structured studies, the Study Designer includes an LLM-assisted study generation workflow. This feature allows a researcher to describe a study in natural language and automatically obtain a draft structured study specification that can then be reviewed and further edited inside the Designer. Rather than replacing manual study design, this functionality is intended to accelerate the creation of initial study drafts and lower the barrier to defining routine EMA protocols.

\begin{figure*}[t]
    \centering

    \begin{subfigure}[t]{0.5\textwidth}
        \centering
        \includegraphics[width=\linewidth,height=0.25\textheight,keepaspectratio]{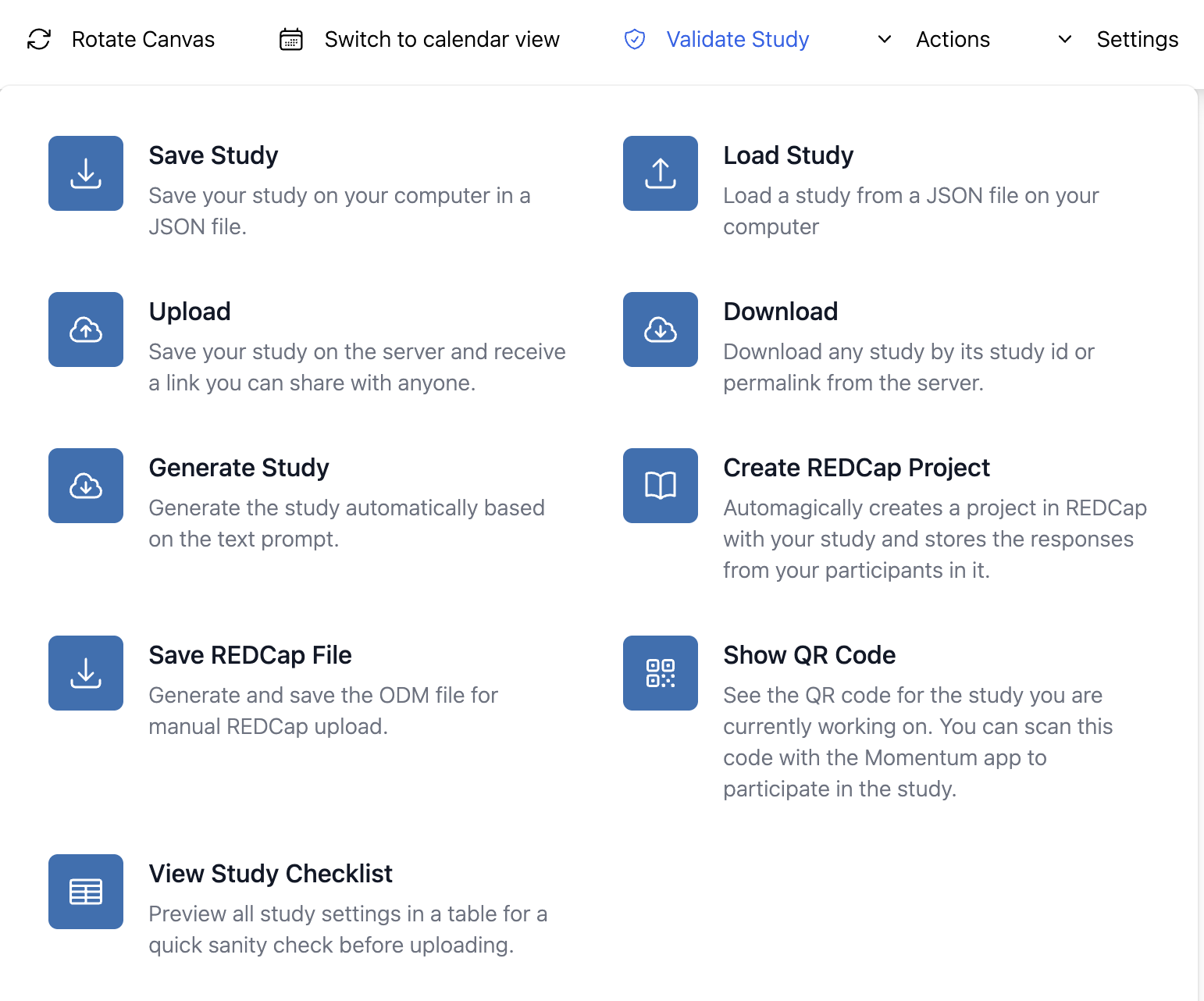}
        \caption{Study Designer action menu with \textit{Generate Study}}
    \end{subfigure}
    \hfill
    \begin{subfigure}[t]{0.4\textwidth}
        \centering
        \includegraphics[width=\linewidth,height=0.25\textheight,keepaspectratio]{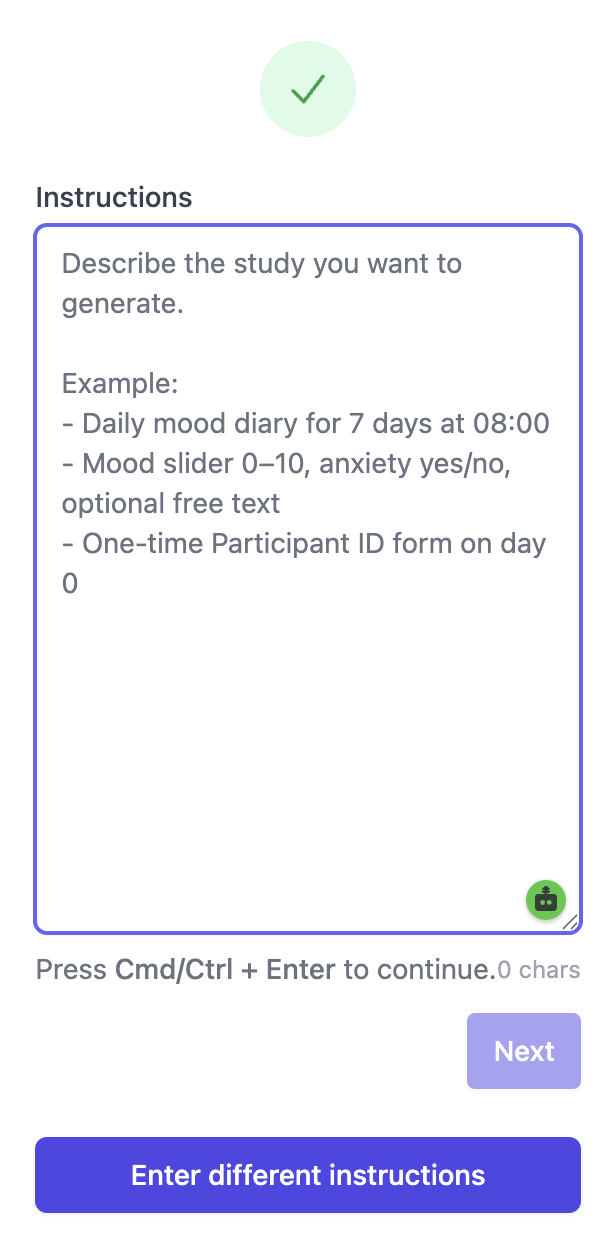}
        \caption{Prompt window for natural-language study description}
    \end{subfigure}

    \caption{LLM-assisted study generation in the Study Designer. A researcher can invoke the generation workflow from the action menu and provide a free-text description of the intended study directly in the Designer interface.}
    \label{fig:llm_ui}
\end{figure*}

The generation workflow is integrated directly into the Designer interface through a dedicated \textit{Generate Study} action. From the action menu, a researcher can open a prompt window and enter a free-text description of the intended study, for example specifying the study duration, scheduling logic, and desired question types (Figure~\ref{fig:llm_ui}). This prompt is sent to a backend generation endpoint, which returns a structured study specification that is immediately compatible with the rest of the platform. Because the generated output follows the same study representation used elsewhere in momenTUM, the result can be inspected, automatically checked for structural validity, modified, exported, and then reviewed by the researcher before deployment through the standard Designer workflow.

A key architectural choice in this component is that the language model does not generate the final study JSON directly. Instead, it first produces an intermediate \textit{study blueprint}: a constrained representation that captures the main study properties, module definitions, schedules, and question specifications in a simplified format. This intermediate structure is then validated against a dedicated blueprint model and compiled into the full study object used by the platform. The compilation step deterministically assigns identifiers, normalizes time formats, resolves defaults, and constructs the complete module, section, and question structure expected by the downstream system. This design reduces the complexity of the generation task and makes the overall pipeline more robust than directly relying on unconstrained model output.

\begin{figure*}[!t]
    \centering
    \includegraphics[width=0.82\linewidth,height=0.38\textheight,keepaspectratio]{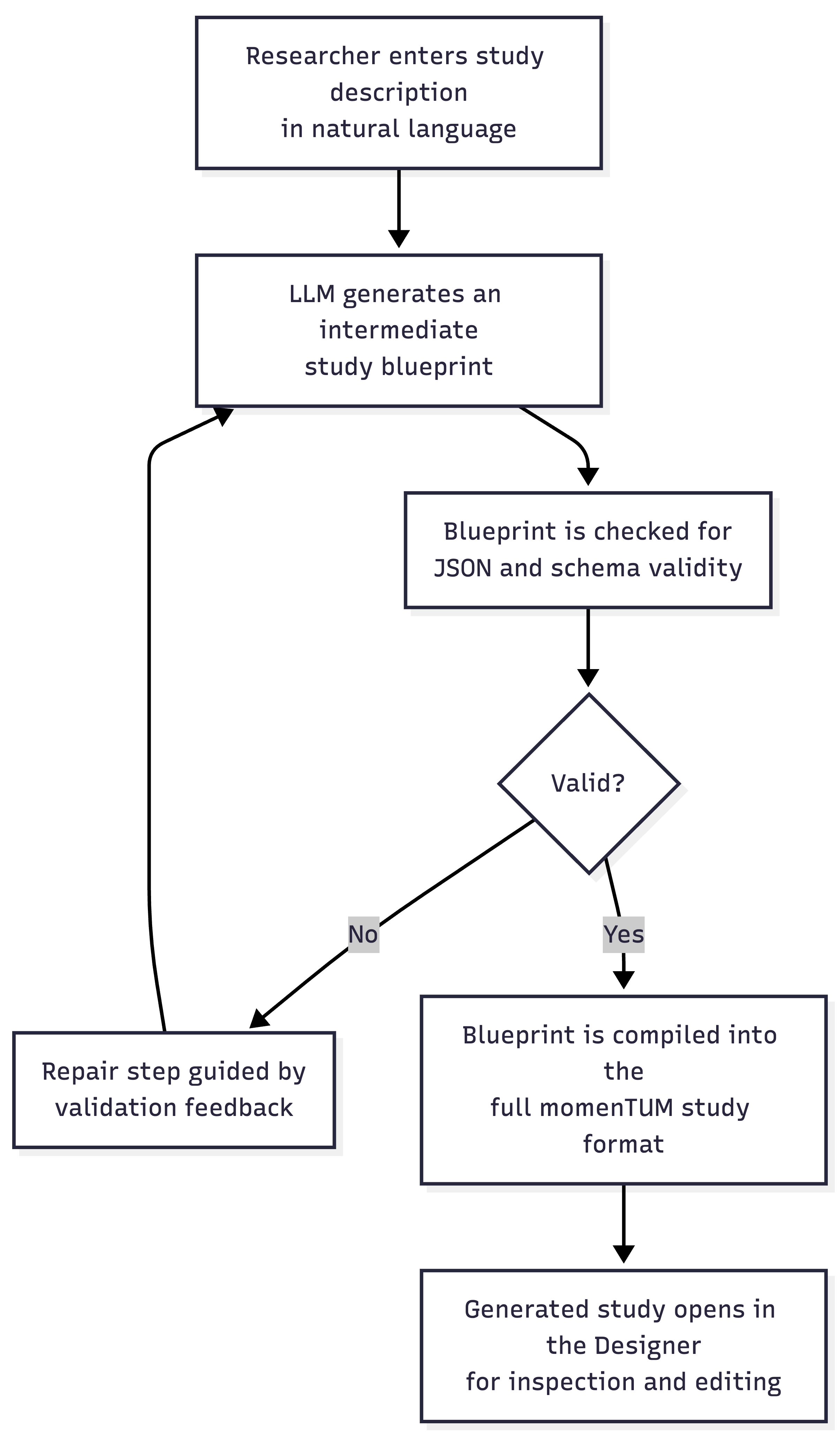}
    \caption{LLM-assisted study generation workflow. A natural-language study description is converted into an intermediate study blueprint, validated and repaired if needed, compiled into the standard momenTUM study format, and returned to the Designer for inspection and further editing.}
    \label{fig:llm_generation_pipeline}
\end{figure*}

The backend generation pipeline also includes iterative repair and validation steps. After the initial model response is received, the output is parsed as JSON and checked against the blueprint schema. If parsing fails or the generated structure does not satisfy the expected constraints, the backend issues a repair step guided by the validation error and schema requirements, and then retries the generation. Once a valid blueprint is obtained, it is compiled into the full study representation and validated again through the platform's standard study model. Figure~\ref{fig:llm_generation_pipeline} summarizes this workflow from natural-language prompt to editable study specification in the Designer.

At its current level of implementation, this feature supports the automatic generation of common study elements including repeated or one-time modules, relative and absolute scheduling patterns, and several standard question types such as sliders, yes/no questions, multiple-choice items, text input, datetime fields, and instruction blocks. This makes it suitable for rapidly drafting many common EMA-style studies while preserving compatibility with the rest of the momenTUM architecture. The automatic checks in this workflow apply to structural correctness within the platform: the generated output must be parseable as JSON, conform to the intermediate blueprint schema, compile into the standard momenTUM study format, and satisfy the corresponding study-model constraints. However, these checks do not establish methodological validity or protocol appropriateness. Generated study specifications therefore remain editable and should still be reviewed by the researcher before deployment, particularly when more specialized study logic or domain-specific constraints are required.

The output of this workflow is a regular momenTUM study specification rather than a separate artifact. This keeps generated drafts editable within the Designer and compatible with the same deployment and data collection pipeline as manually created studies.

\section{Example Deployment Contexts}

momenTUM has been used across several study contexts with different scheduling requirements, questionnaire structures, and durations. These examples are useful for illustrating the range of study designs supported by the platform, although we do not report scientific study outcomes here.

One current use case is AMBIENT-BD \cite{manrai_ambient_2026}, a study on mood, sleep, and circadian rhythms in bipolar disorder conducted at the University of Edinburgh. In this deployment, the platform supports repeated mood and anxiety assessments, a daily morning sleep diary, participant ID entry, and an end-of-period module. The study configuration makes use of relative scheduling, repeated daily prompts across multiple time points, randomization windows, sticky prompts, and response timeouts, illustrating support for intensive repeated psychiatric EMA protocols. The broader AMBIENT-BD study aims to recruit 180 participants, and at the time of reporting 89 participants had consented, with 85 active in the study. The assessment schedule includes four one-week burst EMA periods across an 18-month timeline. At the time of reporting, completed assessments were available for 45 participants in the first EMA period, 28 in the second EMA period, and 4 in the third EMA period, while later periods were still ongoing or had not yet been reached. Between the first and second EMA periods, the protocol was revised to introduce a day delay after registration and to separate the sleep diary from the mood and anxiety questions; the first period included 8 prompts per day and the second period included 9 prompts per day. Across both periods, the median number of completed prompts was 5 per day.

A second example is the EcoSleep cohort study \cite{biller_ecology_2025}, which used earlier versions of the platform in a longitudinal repeated-measurement burst design focused on sleep and circadian research. In this study, healthy adults were followed for 12 months, with three consecutive measurement days every four weeks and four ecological momentary assessments per day during these bursts. This provides an example of long-running repeated field data collection with structured EMA scheduling embedded within a broader multimodal study design.

A further recent deployment is a study on light exposure across the menstrual cycle. This study combines one-time baseline questionnaires with longer daily repeated surveys and an event-triggered menstrual onset module. Its configuration includes two baseline modules administered at enrollment, three daily repeated survey modules scheduled over 40 days, repeated symptom and light exposure questionnaires, daily sleep-related questions, and date-time reporting of menstrual onset. This illustrates the platform’s ability to support mixed study designs that combine baseline, longitudinal diary, and event-based reporting within the same study specification.

\section{Discussion}

momenTUM was developed to address a practical problem in EMA research: the fragmentation of study authoring, participant-facing delivery, backend synchronization, and researcher-side inspection across separate tools. The platform addresses this by treating the structured study specification as the central representation shared across components. In the current system, this shared specification supports study configuration in the Designer, dynamic rendering in the mobile application, synchronization through the backend, and inspection in the researcher dashboard. Taken together, these components provide a coherent workflow for configuring, deploying, and monitoring EMA studies without requiring a separate technical implementation for each study instance.

A main strength of this approach is that study logic is represented explicitly rather than being distributed across application code and platform-specific configuration. This improves consistency between what researchers design, what participants see, and what is later inspected in stored data. It also supports reuse across study periods and related projects, since study behavior can be modified at the level of the specification while preserving the same mobile and backend infrastructure. The deployment examples described in this paper further suggest that this design is practical across different study types, including intensive repeated psychiatric EMA, longitudinal sleep-related protocols, and mixed designs that combine baseline, diary, and event-based reporting.

At the same time, several limitations should be acknowledged. First, this paper focuses on platform design and implementation rather than formal comparative evaluation. We do not report a systematic benchmark against other EMA platforms, nor do we present a dedicated usability study of the Designer, mobile application, or dashboard. Second, although the shared study specification improves consistency, it also makes the quality of the study specification itself especially important: incorrectly configured schedules, options, or identifiers can propagate across multiple parts of the workflow if they are not detected during validation and review. Third, the current platform does not yet provide more advanced participant-specific adaptation features such as individualized timing based on user routines or prior response behaviour. Participant-facing feedback is also limited, apart from configured progress visualizations, and the current system does not include a built-in communication or reporting layer between participants and researchers.

The LLM-assisted study generation workflow should also be interpreted in this context. It is useful as an authoring aid for creating initial drafts of structured studies, but it does not remove the need for researcher review. While the current generation pipeline includes constrained intermediate representations, validation, repair, and compilation into the standard study format, correctness at the schema level is not the same as methodological appropriateness. Study content, scheduling decisions, and deployment details still require domain knowledge and human oversight.

An important future direction is adaptive EMA. The current architecture already separates study content, scheduling, and module logic in a way that could support more personalized and context-sensitive behavior. This could include rule-based adaptation of prompt timing or module availability, as well as more selective assessment strategies that reduce participant burden while preserving informative measurement. Recent work on just-in-time adaptive EMA suggests that adaptive assessment may be useful when EMA is used as a tailoring variable and when burden reduction is important \cite{schneider_just--time_2024}. In the context of momenTUM, this points toward future extensions in which the shared study specification could encode more dynamic triggering and adaptation rules while remaining interpretable and reproducible.

Further development of the platform may include stronger authoring support in the Designer, richer monitoring and visualization in the dashboard, and more adaptive study behavior, including adaptive EMA. Together, these directions would extend the range of study designs and researcher workflows that momenTUM can support.

Overall, momenTUM provides an integrated technical framework for authoring, deploying, synchronizing, and inspecting EMA studies through a shared structured representation. Its main contribution is not a single isolated component, but the way these parts operate together within one reusable study pipeline. This makes the platform suitable for a range of ambulatory assessment designs while keeping studies editable, extensible, and operationally manageable across the full deployment lifecycle.

\section{Conclusion}

We presented momenTUM, a platform for designing, deploying, synchronizing, and monitoring ecological momentary assessment studies through a shared structured study specification. Rather than treating study authoring, participant-facing execution, backend synchronization, REDCap-linked data handling, and researcher-side inspection as separate technical steps, momenTUM connects them through a common machine-readable representation of the study protocol.

In this way, momenTUM turns EMA protocols into reusable, inspectable, machine-readable study objects that can be deployed across studies without rebuilding the mobile application for each new protocol. The deployment examples described in this paper show that this approach is practical across different study contexts, while the LLM-assisted generation workflow illustrates how the same representation can also support earlier stages of study authoring through editable draft study specifications.

\section*{Code Availability}

The source code for momenTUM and its platform components is available through GitHub repositories maintained by the project team at \url{https://github.com/momenTUM-research-platform}.

\section*{Conflict of Interest}

M.S. declares the following potential conflicts of interest in the past five years (2021--2025): academic roles as Member of the Board of Directors of the Society of Light, Rhythms, and Circadian Health (SLRCH), Chair of Joint Technical Committee 20 (JTC20) of the International Commission on Illumination (CIE), Member of the Daylight Academy, and Chair of the Research Data Alliance Working Group Optical Radiation and Visual Experience Data; remunerated roles as Speaker of the Steering Committee of the Daylight Academy, ad-hoc reviewer for the Health and Digital Executive Agency of the European Commission, ad-hoc reviewer for the Swedish Research Council, Associate Editor for \textit{LEUKOS}, examiner for the University of Manchester, Flinders University, and the University of Southern Norway, and consultant for LyS Technologies and RoX Health; research funding and support from the Max Planck Society, Max Planck Foundation, Max Planck Innovation, Technical University of Munich, Wellcome Trust, National Research Foundation Singapore, European Partnership on Metrology, VELUX Foundation, Bayerisch-Tschechische Hochschulagentur (BTHA), BayFrance/Bayerisch-Französisches Hochschulzentrum, BayFOR/Bayerische Forschungsallianz, and Reality Labs Research; honoraria for talks from ISGlobal, the Research Foundation of the City University of New York, and the Stadt Ebersberg, Museum Wald und Umwelt; travel reimbursements from the Daimler und Benz Stiftung; and being named on European Patent Application EP23159999.4A, ``System and method for corneal-plane physiologically-relevant light logging with an application to personalized light interventions related to health and well-being.'' M.S. declares that the disclosed roles and relationships had no influence on the work presented herein. The funders had no role in study design, data collection and analysis, the decision to publish, or preparation of the manuscript.

\section*{Grant Information}

This work was supported by Wellcome [226944, \url{https://doi.org/10.35802/226944}]. 

\textit{The funders had no role in study design, data collection and analysis, decision to publish, or preparation of the manuscript.}

\bibliographystyle{unsrt}  
\bibliography{references}  %%% Remove comment to use the external .bib file (using bibtex).

@article{shiffman_ecological_2008,
	title = {Ecological momentary assessment},
	volume = {10},
	journal = {Applied Clinical Trials},
	author = {Shiffman, Saul and Hufford, Michael},
	month = jan,
	year = {2008},
}

@article{trull_ambulatory_2013,
	title = {Ambulatory assessment},
	volume = {9},
	issn = {1548-5951},
	doi = {10.1146/annurev-clinpsy-050212-185510},
	language = {eng},
	journal = {Annual Review of Clinical Psychology},
	author = {Trull, Timothy J. and Ebner-Priemer, Ulrich},
	year = {2013},
	pages = {151--176},
}

@article{klein_remote_2021,
	title = {Remote {Digital} {Psychiatry} for {Mobile} {Mental} {Health} {Assessment} and {Therapy}: {MindLogger} {Platform} {Development} {Study}},
	volume = {23},
	shorttitle = {Remote {Digital} {Psychiatry} for {Mobile} {Mental} {Health} {Assessment} and {Therapy}},
	url = {https://www.jmir.org/2021/11/e22369},
	doi = {10.2196/22369},
	language = {EN},
	number = {11},
	urldate = {2026-04-22},
	journal = {Journal of Medical Internet Research},
	publisher = {JMIR Publications Inc., Toronto, Canada},
	author = {Klein, Arno and Clucas, Jon and Krishnakumar, Anirudh and Ghosh, Satrajit S. and Auken, Wilhelm Van and Thonet, Benjamin and Sabram, Ihor and Acuna, Nino and Keshavan, Anisha and Rossiter, Henry and Xiao, Yao and Semenuta, Sergey and Badioli, Alessandra and Konishcheva, Kseniia and Abraham, Sanu Ann and Alexander, Lindsay M. and Merikangas, Kathleen R. and Swendsen, Joel and Lindner, Ariel B. and Milham, Michael P.},
	month = nov,
	year = {2021},
	pages = {e22369},
}

@article{mestdagh_m-path_2023,
	title = {m-{Path}: an easy-to-use and highly tailorable platform for ecological momentary assessment and intervention in behavioral research and clinical practice},
	volume = {5},
	issn = {2673-253X},
	shorttitle = {m-{Path}},
	doi = {10.3389/fdgth.2023.1182175},
	language = {eng},
	journal = {Frontiers in Digital Health},
	author = {Mestdagh, Merijn and Verdonck, Stijn and Piot, Maarten and Niemeijer, Koen and Kilani, Ghijs and Tuerlinckx, Francis and Kuppens, Peter and Dejonckheere, Egon},
	year = {2023},
	pages = {1182175},
}

@article{gonzalez-perez_awarns_2023,
	title = {{AwarNS}: {A} framework for developing context-aware reactive mobile applications for health and mental health},
	volume = {141},
	issn = {1532-0464},
	shorttitle = {{AwarNS}},
	url = {https://www.sciencedirect.com/science/article/pii/S1532046423000801},
	doi = {10.1016/j.jbi.2023.104359},
	urldate = {2026-04-22},
	journal = {Journal of Biomedical Informatics},
	author = {González-Pérez, Alberto and Matey-Sanz, Miguel and Granell, Carlos and Díaz-Sanahuja, Laura and Bretón-López, Juana and Casteleyn, Sven},
	month = may,
	year = {2023},
	pages = {104359},
}

@article{shatte_schema_2020,
	title = {schema: an open-source, distributed mobile platform for deploying {mHealth} research tools and interventions},
	volume = {20},
	issn = {1471-2288},
	shorttitle = {schema},
	url = {https://doi.org/10.1186/s12874-020-00973-5},
	doi = {10.1186/s12874-020-00973-5},
	language = {en},
	number = {1},
	urldate = {2026-06-02},
	journal = {BMC Medical Research Methodology},
	author = {Shatte, Adrian B. R. and Teague, Samantha J.},
	month = apr,
	year = {2020},
	pages = {91},
}

@article{schneider_just--time_2024,
	title = {Just-in-time adaptive ecological momentary assessment ({JITA}-{EMA})},
	volume = {56},
	issn = {1554-3528},
	doi = {10.3758/s13428-023-02083-8},
	language = {eng},
	number = {2},
	journal = {Behavior Research Methods},
	author = {Schneider, Stefan and Junghaenel, Doerte U. and Smyth, Joshua M. and Fred Wen, Cheng K. and Stone, Arthur A.},
	month = feb,
	year = {2024},
	pages = {765--783},
}

@article{manrai_ambient_2026,
	title = {Ambient and passive collection of sleep and circadian data in bipolar disorder to understand symptom trajectories and clinical outcomes ({AMBIENT}-{BD}): {Study} protocol [version 2; peer review: 1 approved, 1 approved with reservations]},
	volume = {10},
	doi = {10.12688/wellcomeopenres.23662.2},
	number = {254},
	journal = {Wellcome Open Research},
	author = {Manrai, R and Swiffen, D and Wyse, CA and Gray, D and Ferguson, AC and M. Lopez, L and Campbell, IH and Murray, AL and Caddick, L and Gale, EL and Scorza, LCT and Marwick, KFM and Tsanas, A and Coogan, AN and Spitschan, M and Millar, AJ and Riha, RL and Grigorjeva, MM and Thedie, D and Gardani, M and Zienli?ski, T and Whalley, HC and Smith, DJ},
	year = {2026},
}

@article{biller_ecology_2025,
	title = {The {Ecology} of {Human} {Sleep} ({EcoSleep}) {Cohort} {Study}: {Protocol} for a longitudinal repeated measurement burst design study to assess the relationship between sleep determinants and outcomes under real-world conditions across time of year},
	volume = {34},
	issn = {1365-2869},
	shorttitle = {The {Ecology} of {Human} {Sleep} ({EcoSleep}) {Cohort} {Study}},
	doi = {10.1111/jsr.14225},
	language = {eng},
	number = {2},
	journal = {Journal of Sleep Research},
	author = {Biller, Anna M. and Fatima, Nayab and Hamberger, Chrysanth and Hainke, Laura and Plankl, Verena and Nadeem, Amna and Kramer, Achim and Hecht, Martin and Spitschan, Manuel},
	month = apr,
	year = {2025},
	pages = {e14225},
}
%% and comment out the ``thebibliography'' section.

% %%% Comment out this section when you \bibliography{references} is enabled.

% George Kour and Raid Saabne.
% \newblock Real-time segmentation of on-line handwritten arabic script.
% \newblock In {\em Frontiers in Handwriting Recognition (ICFHR), 2014 14th
%   International Conference on}, pages 417--422. IEEE, 2014.

% George Kour and Raid Saabne.
% \newblock Fast classification of handwritten on-line arabic characters.
% \newblock In {\em Soft Computing and Pattern Recognition (SoCPaR), 2014 6th
%   International Conference of}, pages 312--318. IEEE, 2014.

% \bibitem{hadash2018estimate}
% Guy Hadash, Einat Kermany, Boaz Carmeli, Ofer Lavi, George Kour, and Alon
% \newblock Estimate and replace: A novel approach to integrating deep neural
%   networks with existing applications.
% \newblock {\em arXiv preprint arXiv:1804.09028}, 2018.

% \end{thebibliography}

\end{document}